\documentclass[aps,prl,twocolumn,nofootinbib,superscriptaddress,floatfix]{revtex4-2}

\usepackage{amsmath,amssymb,bm}
\usepackage{graphicx}
\graphicspath{{Figures/}{./}}
\usepackage{booktabs}
\usepackage{microtype}
\usepackage[caption=false]{subfig}
\usepackage{tikz}
\usetikzlibrary{decorations.pathmorphing}
\usepackage[
  pdfborder={0 0 0},
  colorlinks=true,
  citecolor=blue,
  linkcolor=blue,
  urlcolor=blue
]{hyperref}

\usepackage{orcidlink}

\newcommand{\dd}{\mathrm{d}}
\newcommand{\ii}{\mathrm{i}}
\newcommand{\Abs}[1]{\left|#1\right|}
\newcommand{\Order}[1]{\mathcal{O}\!\left(#1\right)}

\newcommand{\kN}{\kappa_N}
\newcommand{\kV}{\kappa_V}
\newcommand{\jzero}{j_0}

\newcommand{\tpar}[2]{\parbox[t]{#1}{\raggedright #2}}
\newcommand{\kallen}{\lambda_{\!K}}
\newcommand{\kallenhalf}[3]{\sqrt{\kallen\!\left(#1,#2,#3\right)}}
\newcommand{\kallenfunc}{\kallen(a,b,c)=a^2+b^2+c^2-2ab-2ac-2bc}

\begin{document}

\title{Spin resummation of heavy quarkonium photoproduction: \\
from the gluonic gravitational form factors to the holographic pomeron}

\author{Kiminad A.~Mamo\,\orcidlink{0000-0003-3123-4398}\,}
\email{ska25005@uconn.edu}
\affiliation{Department of Physics, University of Connecticut, Storrs, CT 06269-3046, USA}

\author{Kemal Tezgin\,\orcidlink{0000-0001-9492-9512}\,}
\email{kemaltezgin@vt.edu}
\affiliation{Department of Physics, Virginia Tech, Blacksburg, VA 24061, USA}

\author{Christian Weiss\,\orcidlink{0000-0003-0296-5802}\,}
\email{weiss@jlab.org}
\affiliation{Theory Center, Jefferson Lab, Newport News, VA 23606, USA}

\date{\today}

\begin{abstract}
Exclusive heavy quarkonium photoproduction probes the proton's gluonic
structure from near-threshold (fixed-spin exchanges, gravitational form factors)
to high energies (reggeized dynamics). We construct a holographic QCD amplitude that
resums the even spin-$j$ gluonic exchanges, with the spin-2 input fixed by lattice QCD GFFs. 
The new framework describes the $J/\psi$ cross section from JLab to HERA energies
in a unified manner. It explains why the spin-2 exchange model for GFF extraction near
threshold is not a controlled approximation and suggests how to improve it.
\end{abstract}

\maketitle

\paragraph{Introduction.} Mapping the proton's \emph{gluonic structure} is a central goal of modern
hadronic physics. The scope of studies now includes the longitudinal momentum distribution
of gluons, the transverse spatial and momentum distributions, and the mechanical properties
carried by the gluon fields. Exclusive heavy quarkonium photoproduction presents a particularly
clean probe of gluonic structure: the heavy quark-antiquark pair couples to the gluon fields,
and its small size provides a short-distance scale. At high energies, heavy quarkonium
production in QCD is described using collinear factorization and generalized parton distributions (GPDs)
\cite{Ryskin:1992ui,Ivanov:2004vd,Collins:1996fb,Radyushkin:1996nd,Diehl:2003ny,Belitsky:2005qn},
or high-energy evolution (BFKL pomeron) and the dipole picture \cite{Kuraev:1977fs,Balitsky:1978ic,%
Golec-Biernat:1998zce,Kowalski:2003hm,Kowalski:2006hc,Watt:2007nr,Bautista:2016xnp},
or the nonperturbative formulation based on gauge-gravity duality (holographic pomeron)
\cite{Brower:2006ea,Brower:2007qh,Ballon-Bayona:2017vlm,Nishio:2014eua,Mamo:2019mka,Mamo:2021tzd,Mamo:2022jhp}.
At energies near threshold, the process is described by the coupling to the proton matrix elements
of the \emph{gluonic energy--momentum tensor}
\cite{Hatta:2018ina,Mamo:2019mka,Hatta:2019lxo,Kharzeev:1995ij,Kharzeev:1998bz,Kharzeev:2021qkd,Guo:2021ibg,Mamo:2022eui,Guo:2023pqw},
the so-called gravitational form factors (GFFs),
which characterize the distributions of mass, spin, and forces inside
the proton \cite{Ji:1996nm,Ji:1996ek,Polyakov:2002yz,Pasquini:2014vua,Burkert:2018bqq,Polyakov:2018zvc,Shanahan:2018nnv,Mamo:2021krl,Cao:2024zlf,Lorce:2025oot,Broniowski:2025ctl,Hechenberger:2025rye,Ji:2025qax}.
The potential for extracting the GFFs from $J/\psi$ production near threshold is demonstrated
by recent results from JLab \cite{GlueX:2019mkq,Duran:2022xag,GlueX:2023pev,007:2026dow,CLAS:2026lls}.

\paragraph{One amplitude, two organizing principles.}
Because heavy quarkonium production follows the same exclusive amplitude from
the near-threshold to the high-energy regime, it exposes a structural tension:
near threshold the amplitude is commonly described by $j=2$ fixed-spin exchanges
(gluonic GFF) \cite{Mamo:2019mka,Guo:2021ibg,Mamo:2022eui,Guo:2023pqw}, while at high energies
it reggeizes and is governed by the leading complex-$j$ singularity
(pomeron) \cite{Brower:2006ea,Brower:2007qh,Ballon-Bayona:2017vlm,Nishio:2014eua,Mamo:2019mka,Mamo:2021tzd}.
Existing approaches typically treat one regime at a time.
In this work we develop a uniform description that interpolates between both regimes,
using the complex angular momentum representation of the amplitude \cite{Gribov:2003nw}
and the dynamics of holographic QCD. New elements relative to earlier holographic studies
\cite{Mamo:2019mka,Mamo:2021tzd} are
the exact evaluation of the full Mellin-Barnes integral, the use of lattice QCD data as
spin-$2$ input, and the implementation of skewness in the complex-$j$ representation.
The fixed-spin and reggeized descriptions correspond to different choices of the integration
contour in the complex $j$-plane with the same analytic integrand. We compare the spin-resummed
and truncated $j =$ 2,4,... amplitudes near threshold. We show that the empirical success of
the spin-$2$ exchange model is not a model-independent consequence of threshold kinematics,
and that the model should be regarded as an effective description, not as the first term of a
controlled finite-spin expansion. This result has important implications
for the extraction of GFFs from heavy quarkonium production near threshold.
In this letter we describe the main steps; details are provided in the Supplemental Material (SM).

\paragraph{Complex $j$ representation.}
The process $\gamma(q)+p(p)\rightarrow V(q')+p(p')$, with
$V=J/\psi$ or $\Upsilon$, is described by the amplitude $\mathcal{M}_V(s,t)$,
which depends on the invariant variables $s=(q+p)^2>0$ and $t=(q-q')^2<0$.
In the complex-$j$ representation the amplitude is expressed as an
even-signature Mellin--Barnes integral \cite{Mamo:2021tzd},
\begin{align}
\mathcal{M}_V(s,t)
&=
-\frac{\pi}{2}\!\int_{\mathcal{C}}
\frac{\dd j}{2\pi\ii}\;
\frac{(s/\kN^2)^j+(-s/\kN^2)^j}{\sin(\pi j)}
\nonumber \\[1.2ex]
&\times
\mathcal{I}_V(j)\;\hat d_j(\eta,t)\,A_g(t,j).
\label{eq:MB}
\end{align}
The universal signature factor involving $(\pm s)^j$  implements the Sommerfeld--Watson transform of the
even-spin exchange sum [SM Eqs.~(\ref{eqS:fixedspin})--(\ref{eqS:factorization})].
The mass scale $\kN$ is the proton mass scale in the soft-wall holographic model and
specified in the following.
$\mathcal{I}_V (j)$ is the $\gamma \rightarrow V$ impact factor, and $A_g(t,j)$ is the proton matrix
element of the spin-$j$ exchange, corresponding to the analytic continuation of the gluon GPD moments;
the explicit form of the functions in the soft-wall holographic model is given in SM
Eqs.~(\ref{eqS:Ij})--(\ref{eqS:T1}). The choice of the integration contour $\mathcal{C}$
in Eq.~(\ref{eq:MB}) has physical implications (fixed-spin vs.\ resummed exchanges)
and is discussed in the following.

%
%
\begin{figure*}[t]
\centering
\includegraphics[width=0.88\textwidth]{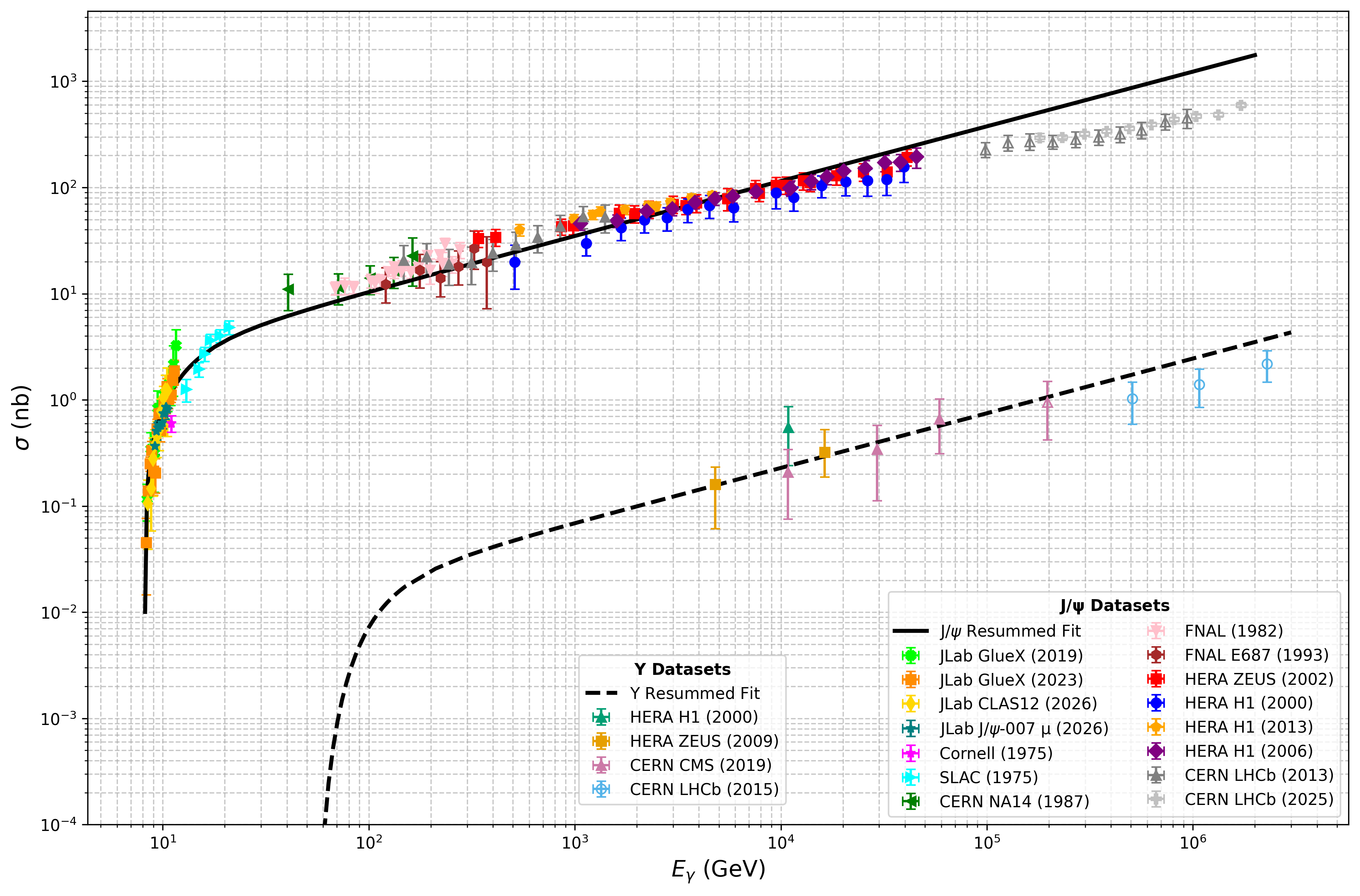}
\caption{Total cross section of exclusive heavy quarkonium photoproduction ($J/\psi, \Upsilon$)
versus the photon lab energy $E_\gamma$. \textit{Solid curve:} $J/\psi$ cross section
in the holographic QCD model based on the complex-$j$ representation, Eqs.~(\ref{eq:MB})--(\ref{eq:dhat}).
Parameters and fit procedure are described in the text. The fit includes the data from
the fixed-target experiments (JLab, Cornell, SLAC, CERN, FNAL) 
\cite{GlueX:2019mkq,GlueX:2023pev,007:2026dow,CLAS:2026lls,%
Gittelman:1975ix,Camerini:1975cy,Barate:1986fq,Binkley:1981kv,E687:1993hlm}
and the HERA collider \cite{H1:2000kis,ZEUS:2002wfj,H1:2005dtp,H1:2013okq};
it does not include the highest-energy data from CERN LHCb ($E_\gamma > 10^5$ GeV) \cite{Aaij:2013jxj,LHCb:2024pcz}.
\textit{Dashed curve:} $\Upsilon$ cross section in the same holographic QCD model.
The fit uses the holographic exchange parameters from the $J/\psi$ analysis and infers only the
normalization from the HERA data \cite{H1:2000kis,ZEUS:2009asc}; it does not include the
CERN data \cite{LHCb:2015wlx,CMS:2018bbk}.}
\label{fig:global}
\end{figure*}

\paragraph{BPST cut and strong-coupling input.}
Strong-coupling reggeization enters the complex-$j$ representation through the
BPST spin-dimension relation \cite{Brower:2006ea,Brower:2007qh},
\begin{equation}
\Delta(j)=2+\sqrt{2\sqrt{\lambda}\,\bigl(j-\jzero\bigr)},
\qquad
\jzero=2-\frac{2}{\sqrt{\lambda}},
\label{eq:Delta}
\end{equation}
where $\lambda$ is the 't Hooft coupling.
It produces a square-root branch point at $j=\jzero$ and a cut for $j\le \jzero$
in the function $\mathcal{I}_V(j)$ [SM Eq.~(\ref{eqS:Delta})]. 
The large-$s$ behavior of the integral Eq.~\eqref{eq:MB} is controlled by the
BPST singularity. For $s$ near threshold, a fixed pole at $j = 2$ with an effective residue
provides a phenomenological model of the amplitude; below we compare it with the full
spin-resummed result.

\paragraph{Skewness, polynomiality, and $D$-term.}
The GPD moments have the polynomiality property following from relativistic
covariance, which dictates their dependence on the skewness variable describing the
longitudinal momentum transfer to the proton \cite{Polyakov:1999gs,Diehl:2003ny,Belitsky:2005qn}.
In Eq.~\eqref{eq:MB} we use the holographic implementation of this property through the factor
\begin{align}
& \hat d_j(\eta,t) =
{}_2F_{1}\!\left(
-\frac{j}{2},\frac{1-j}{2};\frac{1}{2}-j;
z_S \right),
\nonumber\\
& z_S \equiv -\frac{4m_N^2\eta^2}{t-m_S^2},
\hspace{1em}
\eta \equiv
\frac{M_{V}^2}{2(s-m_N^2)-M_{V}^2+t}.
\label{eq:dhat}
\end{align}
Here $\eta$ is the skewness variable for heavy quarkonium photoproduction, and $M_V$ is
the quarkonium mass. Eq.~(\ref{eq:dhat}) is obtained by extending the polynomiality
condition of the GPD moments to complex $j$. For positive even $j$, Eq.~\eqref{eq:dhat}
terminates to a polynomial in $\eta^2$. For $j = 2$ it reproduces the well-known
expression for the $D$-term describing the highest power $\propto \eta^2$ in the second moment
of the gluon GPD [SM Eqs.~(\ref{eqS:Aj_plus_D})--(\ref{D2})] \cite{Polyakov:1999gs,Diehl:2003ny,Belitsky:2005qn}.
The parameter $m_S$ is fixed from the $t$-dependence of the gluonic $D$-term form factor.
For physical $t<0$, the argument $z_S$ is positive, and the function $\hat{d}_j$ suppresses
contributions of large real $j \rightarrow \infty$
[SM Eqs.~(\ref{eqS:zS})--(\ref{eqS:Fjsmallz})].

\paragraph{Contour and numerical evaluation.}
The complex-$j$ integral Eq.~\eqref{eq:MB} is originally defined with the Sommerfeld-Watson contour,
which encloses the poles of the signature factor at real positive $j=2,4,\ldots$; this contour
represents the amplitude as a sum over spins and can be used also for the fixed-spin truncation.
For evaluating the reggeized amplitude numerically we choose the Bromwich contour
$\mathcal{C}: j=c+\ii\nu$, with $\jzero<c<2$ and $-\infty < \nu < \infty$, a vertical line
to the right of the BPST branch point and to the left of the first fixed-spin pole.
The contour is then deformed leftward toward the BPST cut; the large left arc vanishes,
while a right-closed large semicircle is not part of the numerical prescription.
Details of the contour prescription and convergence are given in the SM
[SM Fig.~\ref{fig:contour_app_MB},
Eqs.~(\ref{eqS:Phi})--(\ref{eqS:Phi_bound}),
and Figs.~\ref{fig:MB_decay_im_app}--\ref{fig:MB_arc_app}].
Kinematic poles from the skewness factor and the $j=0,-2,\ldots$ signature/$\Gamma$-function
singularities crossed in auxiliary deformations are treated by the standard indentation/residue
prescription.

\paragraph{Differential and total cross section.}
The unpolarized differential cross section is
\begin{equation}
\frac{\dd\sigma}{\dd t}
=
\frac{\Abs{\mathcal{M}_V(s,t)}^2}{16\pi\,(s-m_N^2)^2}.
\label{eq:dsdt}
\end{equation}
The total cross section is obtained by integrating over the physical $t$ range,
\begin{equation}
\sigma(s)=\int_{t_{\max}(s)}^{t_{\min}(s)}\dd t\,
\frac{\dd\sigma}{\dd t};
\label{eq:sigma_total_main}
\end{equation}
the kinematic bounds are given in SM Eq.~(\ref{t_bounds}); note that $t_{\max}\le t_{\min}\le0$.
For $s$ near threshold, $|t_{\rm min}|$ is large and $t_{\rm min} - t_{\rm max}$ is small;
for $s \rightarrow \infty$, $t_{\rm min} \rightarrow 0$ and $t_{\rm max} \rightarrow -\infty$,
and the $t$-integral is dominated by the region of small $t$.

\paragraph{Global fit of $J/\psi$ total cross section data.}
Using the complex-$j$ representation of the amplitude and the holographic model input,
we perform a global fit of the experimental data on the total cross section of exclusive $J/\psi$
photoproduction from JLab
to HERA energies \cite{GlueX:2019mkq,GlueX:2023pev,007:2026dow,CLAS:2026lls,Gittelman:1975ix,%
Binkley:1981kv,Barate:1986fq,H1:2000kis,ZEUS:2002wfj,H1:2005dtp,H1:2013okq}.
The proton spin-$2$ GFF input is fixed independently by lattice QCD results
in the $\overline{\mathrm{MS}}$ scheme at
$\mu=2~\mathrm{GeV}$ \cite{Pefkou:2021fni} [SM Fig.~\ref{fig_A-D_latticefit}].
The remaining parameter determination is summarized in SM Table~\ref{tab:params}.
Only the coupling $\lambda$ and the overall normalization of the amplitude
(contained in the impact factor $\mathcal{I}_V$)
are fitted to the $J/\psi$ photoproduction data, giving
\begin{equation}
\lambda\simeq 8.13
\label{lambda_fit}
\end{equation}
[SM Eq.~(\ref{eqS:chi2_global}), (\ref{fit_results})]. Figure~\ref{fig:global} shows the resulting fit
of the cross section data as a function of the equivalent photon lab energy,
$E_\gamma \equiv (s - m_N^2)/2 m_N$. A very satisfactory uniform description is achieved
in this minimal framework. 
The model describes both the near-threshold and the high-energy regime, clearly showing the
power of the $j$-resummation. (The theory-based fit performed here does not attempt to account
for inconsistencies of data from different experiments in the near-threshold region, which is reflected
in the $\chi^2$ value; see SM for discussion of possible refinements.)

\paragraph{Strong coupling from energy dependence.}
The coupling $\lambda$ controls the
effective Regge behavior of the amplitude through the BPST cut, Eq.~(\ref{eq:Delta}).
Equation~(\ref{lambda_fit}) represents the effective value obtained from the fit
to the full JLab-to-HERA data set, in which the near-threshold data have a major impact.
An earlier fit to the HERA data alone \cite{Mamo:2021tzd}
obtained a larger coupling, $\lambda=11.243$. In that analysis the Mellin-Barnes integral
was approximated by the contribution of the BPST cut in the high-energy regime, whereas
here it is evaluated numerically, and the fit includes the full threshold-to-HERA data set.

%
%
\begin{figure}[t]
  \centering
  \includegraphics[width=0.9\columnwidth]{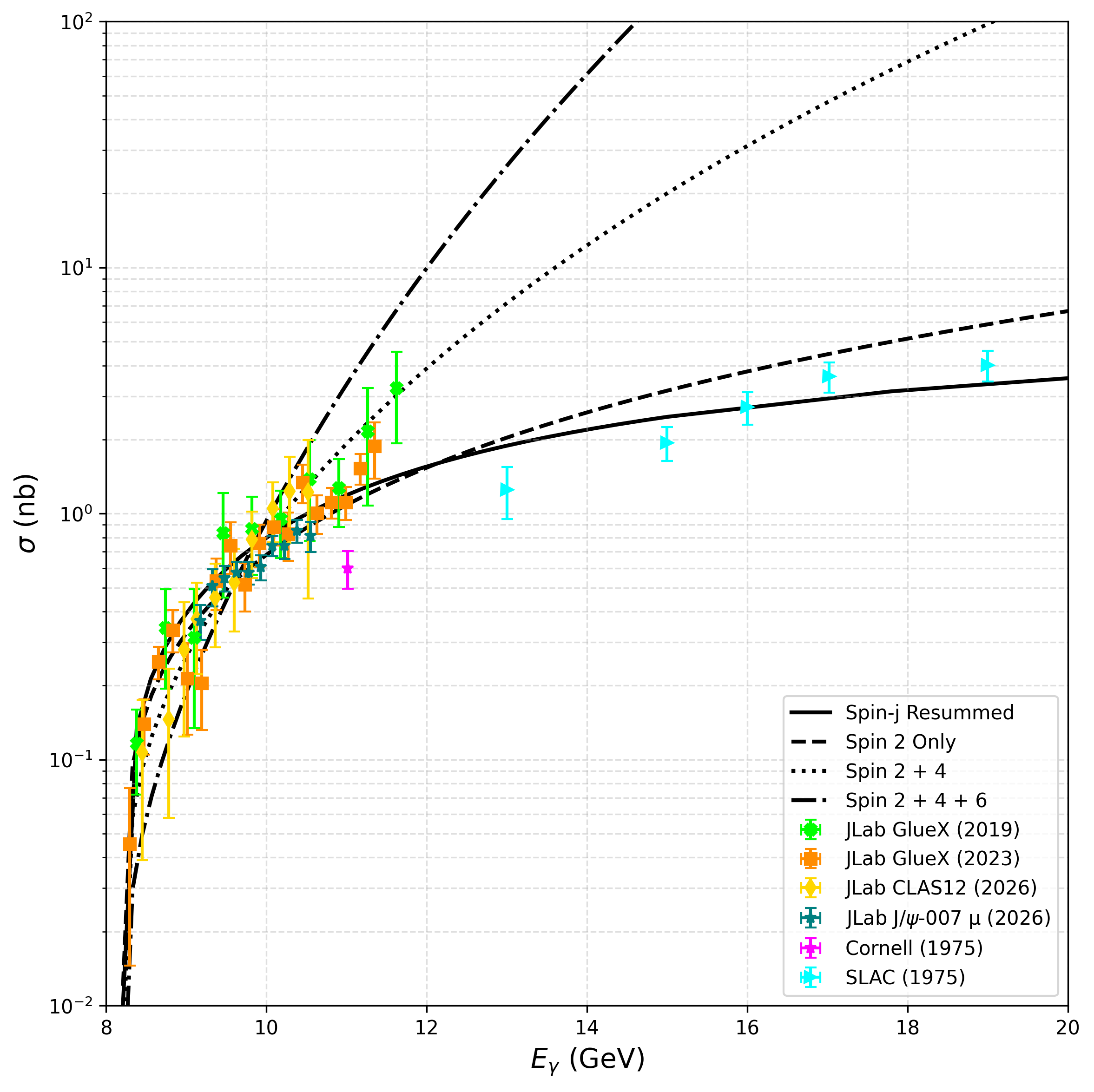}
\caption{Resummed vs.\ fixed-spin models of $J/\psi$ photoproduction at near-threshold energies.
\textit{Solid line:} Resummed spin exchanges evaluated with the Mellin-Barnes integral, Eq.~\eqref{eq:MB}.
The holographic model parameters are determined by fit including high-energy data (see text).
\textit{Dashed, dotted, and dashed-dotted lines:}
Fixed-spin exchange models with $j = 2, 2 + 4, 2 + 4 + 6$, obtained from the even-spin representation
summarized in SM Eqs.~(\ref{eqS:fixedspin})--(\ref{eqS:factorization}).
The normalization is determined by fitting the JLab GlueX data with the respective truncation.}
\label{fig:trunc}
\end{figure}
\paragraph{Fixed-spin vs.\ resummed exchanges.}
Using our framework we can now compare the $j$-resummed model with fixed-spin exchange models
at near-threshold energies. Figure~\ref{fig:trunc} shows the resummed model, with $\lambda$
given by Eq.~\eqref{lambda_fit}, and the fixed-spin models with $j=2$, $2+4$, and $2+4+6$.
For each fixed-spin model the normalization is separately fitted to the near-threshold JLab GlueX data;
the curves are therefore not successive terms with a common normalization.
The fixed-spin models exhibit high-energy behavior as
$\mathcal{M}_V\sim s^2,s^4,s^6$, so the higher-spin truncations rapidly overshoot the data.
The resummed model instead exhibits a Regge-like growth controlled by the BPST cut,
with $\mathcal{M}_V\sim s^{\jzero}$ and
\begin{equation}
\jzero=2-\frac{2}{\sqrt{\lambda}}\simeq1.30\qquad(\lambda\simeq8.13).
\end{equation}
One observes that the energy dependence of the $j = 2$ model is similar to that
of the resummed model up to $E_\gamma \approx$ 12 GeV, simply because the value of $j_0$
is close to 2. However, this should not be regarded as evidence of a controlled expansion
in spin, as the addition of higher terms $j = 4, 6, \ldots$ does not lead to an improved
approximation. More basically, there is no model-independent connection between
the strength of the BPST-cut and the residue of the $j=2$ pole in this representation.

\paragraph{Implications for GFF extraction.} Our findings have implications
for the extraction of the spin-2 gluonic GFFs from near-threshold $J/\psi$ data.
The $j=2$ exchange model should be regarded as an effective model of the amplitude,
not as the first term of a controlled expansion in the spin of the exchanges.
Corrections to the GFF extraction with the spin-2 exchange model should be estimated
with the resummed-$j$ amplitude, not by adding individual $j=4,6,\ldots$ exchanges.
This does not invalidate the spin-$2$ exchange model, but it limits what can be inferred
from its phenomenological success.
These conclusions can be extended to the GPD-based
formulation of heavy quarkonium photoproduction near threshold, where the
large-skewness expansion is used to establish an approximate relation of the amplitude
with the second moment of the GPDs \cite{Guo:2021ibg,Guo:2023pqw}.
The large-skewness expansion in the GPD formulation is structurally analogous
to the fixed-spin expansion in the holographic model. It can be shown that corrections
to the second-moment approximation involve the full set of higher moments of the GPDs
and cannot be organized by individual moments. The same applies to exclusive
$\phi$ meson electroproduction near threshold \cite{Hatta:2021can,Hatta:2025vhs}.
The detailed arguments will be presented elsewhere.

\paragraph{Validation with $J/\psi$ differential cross section data.}
Further validation of the holographic model results can be obtained by comparing with the
$t$-dependent differential cross section data. With the coupling $\lambda$ and
the impact factor normalization fixed by the total cross section fit, the model
predicts the $t$-dependence of the differential cross section without further parameters.
In SM Fig.~\ref{fig:diff_cs_jlab} we compare this prediction with representative
near-threshold differential cross section bins from JLab GlueX and CLAS
\cite{GlueX:2023pev,CLAS:2026lls}. (No differential cross-section data are used in the
$\chi^2$ minimization, and no additional $t$-slope or bin-dependent normalization parameters are introduced.)
The agreement provides a strong validation of the $j$-resummed holographic dynamics,
the lattice QCD proton-side input, and the normalization obtained from the total cross section fit.
A follow-up study will analyze all differential cross section data from JLab to HERA.

\paragraph{From $J/\psi$ to $\Upsilon$.}
The holographic model with $j$ resummation describes exclusive $\Upsilon$ photoproduction in the
same unified framework. Only the impact factor changes compared to the $J/\psi$;
the $j$ exchanges and their couplings to the proton remain the same. 
Figure~\ref{fig:global} shows the holographic prediction for the $\Upsilon$ cross section.
Here the model input is the same as in the $J/\psi$ analysis, and $\lambda$ is
taken as determined by the fit to the $J/\psi$ data, Eq.~\eqref{lambda_fit};
only the overall normalization
(contained in the impact factor) is fitted to the HERA $\Upsilon$ data \cite{H1:2000kis,ZEUS:2009asc}.
The $\Upsilon$ comparison therefore does not test a parameter-free prediction; rather,
it tests whether the dynamics with the same exchanges and proton couplings as in the $J/\psi$
can accommodate the $\Upsilon$ data once the new channel normalization is specified.
A good description of the energy dependence of the HERA $\Upsilon$ data is achieved.
[The $j$-dependence of the holographic impact factor $\mathcal{I}_V(j)$,
SM Eqs.~(\ref{eqS:Ij}) and (\ref{eqS:Cj}), depends on the parameter $\kappa_V$
representing the scale of the produced heavy meson, which leads to a dependence of the
cross section on the external meson mass. In the present test we set $\kappa_{\Upsilon} =
\kappa_{J/\psi}$, so that this effect is not included; it easily could be included in a
more comprehensive analysis.] 
The holographic predictions for the energy dependence near threshold can be tested when
near-threshold $\Upsilon$ data become available at the Electron-Ion
Collider (EIC) \cite{Joosten:2018gyo,Gryniuk:2020mlh}.

\paragraph{Onset of unitarity corrections at ultra-high energies.}
The single-exchange resummation in Eq.~\eqref{eq:MB} (one BPST Regge cut) implies an effective
powerlike rise, governed by the leading $j$-plane singularity, and therefore cannot remain valid
indefinitely at asymptotically large $s$.
Consistent with this expectation, the highest-energy $J/\psi$ data in Fig.~\ref{fig:global}
(LHCb, $E_\gamma \gtrsim 10^{5}\,\mathrm{GeV}$) fall below the single-BPST curve, suggesting that
multi-Pomeron exchange and eikonalization/unitarization effects become important in this regime
\cite{Brower:2007xg,Cornalba:2007fs}.
Such corrections are precisely the mechanism by which Regge growth is tamed, and in hadronic
scattering they underlie the $\ln^{2}s$ behavior compatible with the Froissart--Martin bound
\cite{Froissart:1961ux,Martin:1962rt,Mamo:2025hur}.

\paragraph{Summary.}
We have constructed a framework for the analysis of heavy quarkonium photoproduction based on
the complex-$j$ representation of the amplitude and dynamical input from holographic QCD.
It gives a single resummed amplitude connecting the threshold spin-$2$/GFF organization
with the high-energy BPST-pomeron organization.
The successful fit of the $J/\psi$ total cross section data from JLab to HERA
demonstrates the phenomenological role of the resummation, with additional validation from
the $t$-dependent differential cross sections and the $\Upsilon$ data.
The comparison of the resummed and fixed-spin exchanges near threshold
shows that the spin-$2$ exchange is an effective model, not a controlled approximation
to the resummed amplitude. The framework can be used for next-generation studies of the proton's
gluonic structure in heavy quarkonium photoproduction at JLab, EIC and other facilities.
It can be extended to heavy quarkonium electroproduction ($Q^2$ dependence, $L$ and $T$ amplitudes) and other
exclusive meson production channels (pseudoscalar and vector, light quark states).

\begin{acknowledgments}
This material is based upon work supported by the U.S.~Department of Energy, Office of Science,
Office of Nuclear Physics under 
Contract No.\ 89243126CSC000213.
The work of K.A.M.\ was supported also by the U.S.~National Science Foundation under Grant No.\ 2412625, and by the U.S.~Department of Energy,
Office of Science, Office of Nuclear Physics 
under the umbrella of the Quark-Gluon Tomography
(QGT) Topical Collaboration under Contract DE-SC0023646. The work of K.T. was supported by the
U.S.~Department of Energy,
Office of Science, Office of Nuclear Physics 
through the EXCLAIM collaboration 
under 
Contracts No.\ DE-SC0016286 and DE-SC0024644.

\end{acknowledgments}

\bibliography{references}

\clearpage
\onecolumngrid

\setcounter{secnumdepth}{4}
\setcounter{tocdepth}{2}
\setcounter{section}{0}
\setcounter{subsection}{0}
\setcounter{subsubsection}{0}
\setcounter{paragraph}{0}
\renewcommand{\thesection}{\Roman{section}}
\renewcommand{\thesubsection}{\Alph{subsection}}
\renewcommand{\thesubsubsection}{\arabic{subsubsection}}
\renewcommand{\theparagraph}{\alph{paragraph}}

\setcounter{equation}{0}
\setcounter{figure}{0}
\setcounter{table}{0}
\renewcommand{\theequation}{S\arabic{equation}}
\renewcommand{\thefigure}{S\arabic{figure}}
\renewcommand{\thetable}{S\arabic{table}}

\renewcommand{\theHequation}{S\arabic{equation}}
\renewcommand{\theHfigure}{S\arabic{figure}}
\renewcommand{\theHtable}{S\arabic{table}}
\renewcommand{\theHsection}{S.\Roman{section}}
\renewcommand{\theHsubsection}{S.\Roman{section}.\Alph{subsection}}
\renewcommand{\theHsubsubsection}{S.\Roman{section}.\Alph{subsection}.\arabic{subsubsection}}
\renewcommand{\theHparagraph}{S.\Roman{section}.\Alph{subsection}.\arabic{subsubsection}.\alph{paragraph}}

\begin{center}
  {\Large\bfseries Supplemental Material:\par}
  \vspace{0.35em}
  {\Large\bfseries Spin resummation of heavy quarkonium photoproduction:\par}
  {\Large\bfseries from the gluonic gravitational form factors to the holographic pomeron\par}
  \vspace{1.0em}
  {\normalsize Kiminad A.~Mamo, Kemal Tezgin, and Christian Weiss\par}
  \vspace{0.55em}
  {\normalsize (Dated: \today)\par}
\end{center}
\vspace{0.6em}

\begingroup
\makeatletter
\newif\ifsm@numberedtocentry
\newcommand{\sm@checknumberline}[1]{%
  \def\sm@tmp{#1}%
  \ifx\sm@tmp\@empty\else\global\sm@numberedtocentrytrue\fi
}
\newcommand{\sm@onlynumberedtoc}[3]{%
  \global\sm@numberedtocentryfalse
  \begingroup
    \let\numberline\sm@checknumberline
    \setbox\z@=\hbox{#2}%
  \endgroup
  \ifsm@numberedtocentry
    #1{#2}{#3}%
  \fi
}
\let\sm@original@section\l@section
\renewcommand{\l@section}[2]{\sm@onlynumberedtoc\sm@original@section{#1}{#2}}
\tableofcontents
\makeatother
\endgroup

\section{Kinematics and cross section}
\label{secS:kinematics}
In this section we describe the kinematic variables and cross section of the
process of heavy quarkonium photoproduction on the proton,
\[
\gamma(q)+p(p)\to V(q')+p(p').
\qquad V=J/\psi,\Upsilon ,
\]
The photon is assumed to be real, $Q^2 \equiv -q^2 = 0$; the hadron 4-momenta satisfy
$p^2 = p^{\prime 2} = m_N^2$ (nucleon mass) and $q^{\prime 2} = M_V^2$ (heavy quarkonium mass).
We define the auxiliary 4-momenta
\begin{align}
& \tilde{p}=p+p',
\qquad
\tilde{q}=\frac{1}{2}\left(q'+q\right),
\qquad
\Delta \equiv q - q' = p' - p,
\end{align}
and the invariant variables
\begin{align}
&s=(p+q)^2=W^2 > 0,
\qquad
t = \Delta^2<0 .
\end{align}

In the $\gamma p$ center-of-mass (CM) frame,
\begin{alignat}{2}
& |\bm{q}_{\rm CM}|
=\frac{s-m_N^2}{2\sqrt{s}},
&& E_{\rm CM} = |\bm{q}_{\rm CM}|,
\\
& |\bm{q}'_{\rm CM}|
=\frac{\kallenhalf{s}{M_V^2}{m_N^2}}{2\sqrt{s}},
\hspace{1em}
&& E'_{\rm CM} = \frac{s+M_V^2-m_N^2}{2\sqrt{s}},
\end{alignat}
where
\begin{align}
\kallenfunc .
\end{align}
In terms of the CM scattering angle $\theta$, the invariant momentum transfer is
\begin{equation}
t = M_V^2 - 2 E_{\rm CM} E'_{\rm CM} + 2|\bm{q}_{\rm CM}| |\bm{q}'_{\rm CM}| \cos\theta .
\end{equation}
The kinematic limits are given by
\begin{equation}
t_{\min}(s)\equiv t(\theta=0),
\qquad
t_{\max}(s)\equiv t(\theta=\pi),
\qquad
t_{\max}\le t_{\min}\le 0,
\label{t_bounds}
\end{equation}
where $t_{\min} (t_{\rm max})$ corresponds to forward (backward) scattering.
At the threshold energy $\sqrt{s}_{\rm th} \equiv M_V + m_N$ the CM momentum in the final state
is zero $|\bm{q}'| = 0$, and the phase space collapses, $|t_{\rm max}(s_{\rm th})|
= |t_{\rm min}(s_{\rm th})|$.

The differential cross section of the unpolarized photoproduction process is
\begin{equation}
\frac{\dd\sigma}{\dd t}
=
\frac{\Abs{\mathcal{M}_V(s,t)}^2}{16\pi\,(s-m_N^2)^2}.
\label{eqS:dsdt}
\end{equation}
where $\mathcal{M}_V$ is the scattering amplitude with the conventional relativistic normalization
of states. The total cross section is then
\begin{equation}
\sigma(s)=\int_{t_{\max}(s)}^{t_{\min}(s)}\dd t\;\frac{\dd\sigma}{\dd t}.
\label{eqS:sigma_total}
\end{equation}

In the data analysis in Sec.~\ref{secS:globalfit}  and in Fig.~\ref{fig:global} of the main text
we use as energy variable the equivalent photon lab energy, denoted by $E_\gamma$, 
\begin{equation}
E_\gamma=\frac{s-m_N^2}{2m_N}.
\end{equation}
In this variable the threshold is at
\begin{equation}
E_{\gamma, {\rm th}} = \textrm{8.21 GeV ($J/\psi$)}, \hspace{1em} \textrm{57.2 GeV ($\Upsilon$)}
\end{equation}

\section{Angular momentum resummation}
\label{secS:MBderivation}

\subsection{Mellin--Barnes integral}
In this section we derive the Mellin-Barnes representation of the heavy quarkonium
photoproduction amplitude as the Sommerfeld-Watson transform of the discrete sum
of even-spin exchanges.

\begin{figure}[t]
\centering
\resizebox{0.7\linewidth}{!}{%
\begin{tikzpicture}[x=1.05cm,y=1.05cm]
  \draw[->,thick] (-3.0,0) -- (5.2,0) node[below] {$\mathrm{Re}(j)$};
  \draw[->,thick] (0,-2.5) -- (0,2.5) node[left] {$\mathrm{Im}(j)$};

  \fill (1.3,0) circle (1.6pt);
  \node[below] at (1.3,0) {$j_0$};
  \draw[decorate,decoration={zigzag,segment length=5pt,amplitude=1.2pt},thick]
    (-3.0,0) -- (1.3,0);
  \node[fill=white,inner sep=1.2pt] at (-1.4,0.25) {\scriptsize BPST cut};

  \fill (0.5,0) circle (1.4pt);
  \node[below] at (0.5,-0.05) {\scriptsize $1/2$};
  \fill (1.5,0) circle (1.4pt);
  \node[above] at (1.35,0.01) {\scriptsize $3/2$};
  \node[align=center,fill=white,inner sep=1.2pt] at (0.95,-0.65)
  {\scriptsize apparent\\[-1pt]\scriptsize skewness poles};

  \fill (2.0,0) circle (1.5pt);
  \node[below] at (2.0,0) {\scriptsize $2$};
  \fill (4.0,0) circle (1.5pt);
  \node[below] at (4.0,0) {\scriptsize $4$};
  \node[above] at (3.0,0.05) {\scriptsize even-spin poles};
  \draw[gray,dashed,rounded corners] (1.75,-0.38) rectangle (4.25,0.38);
  \node[gray,above] at (3.5,0.52) {\scriptsize formal fixed-spin SW contour};

  \draw[thick,blue,->] (1.6,-2.2) -- (1.6,2.2);
  \node[blue,right] at (1.6,2.05) {\scriptsize $\mathcal{C}_B:\;j=c+\ii\nu,\;j_0<c<2$};

  \draw[thick] (1.6,2.2) arc[start angle=90,end angle=270,radius=2.2];
  \node[left] at (-0.45,-0.75) {\scriptsize left deformation};

  \node[align=left,fill=white,inner sep=2pt] at (-2.10,-1.65)
  {\scriptsize deformation toward BPST cut\\[-1pt]
   \scriptsize large left arc $\to 0$};
\end{tikzpicture}}
\caption{Illustrative $j$-plane geometry.  The fixed-spin Sommerfeld--Watson representation
encircles the positive even-spin poles $j=2,4,\ldots$.  The numerical Reggeized amplitude is
evaluated on the upward Bromwich line $j=c+\ii\nu$, with $j_0<c<2$.  The convergence-relevant
deformation is to the left, toward the BPST cut; isolated half-integer skewness poles are handled
by the indentation/residue prescription and do not determine the leading high-energy behavior.}
\label{fig:contour_app_MB}
\end{figure}
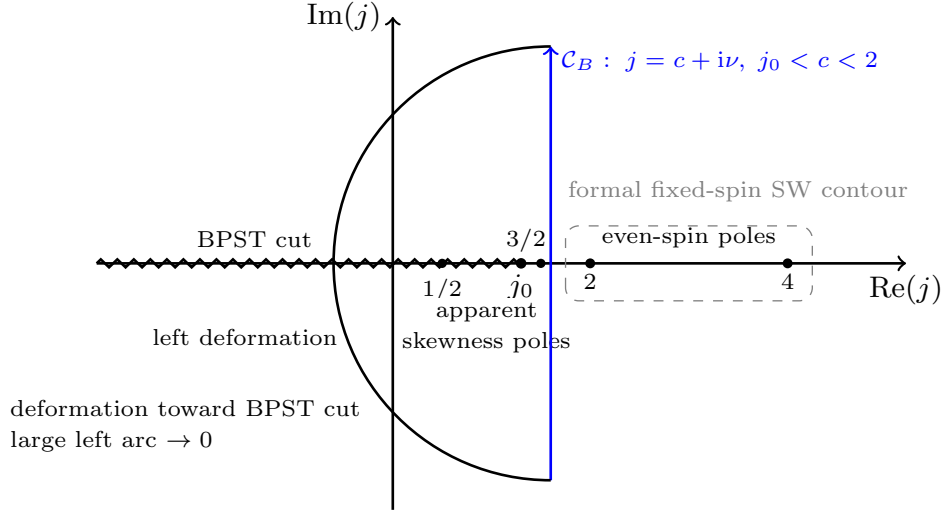

At fixed even spin $j=2,4,6,\ldots$, the $t$-channel exchange contribution has the
schematic structure
\begin{equation}
\mathcal{M}_{V,j}(s,t)=\beta_{V,j}(t)\left(\frac{s}{\kN^2}\right)^{\!j},
\label{eqS:fixedspin}
\end{equation}
where the residue $\beta_{V,j}(t)$ describes the coupling of the spin-$j$ exchange to
the external systems.  The even-signature amplitude is
\begin{equation}
\mathcal{M}_V(s,t)=\sum_{j\in 2\mathbb{N}}\mathcal{M}_{V,j}(s,t),
\qquad
2\mathbb{N}\equiv\{2,4,6,\ldots\}.
\label{eqS:even_sum}
\end{equation}
For a finite truncation with maximal retained spin $J$,
\begin{equation}
\mathcal{M}_V^{(J)}(s,t)
=
\sum_{j=2,4,\ldots,J}
\beta_{V,j}(t)\left(\frac{s}{\kN^2}\right)^{\!j}
\underset{s/\kN^2\gg1}{\longrightarrow}
\beta_{V,J}(t)\left(\frac{s}{\kN^2}\right)^{\!J},
\label{eqS:finite_truncation}
\end{equation}
provided the highest-spin residue is nonzero. Thus every finite fixed-spin model in this
holographic QCD construction is nonuniform in energy and is eventually dominated by its highest
retained power. Adding $j=4,6,\ldots$ does not define a systematically improving global
approximation. The separately normalized curves in main-text Fig.~\ref{fig:trunc} are
therefore effective fixed-spin models, not successive common-normalization partial sums.
The $j=0$ pole is not included in the fixed-spin sum.  Assuming analytic continuation of
$\beta_{V,j}(t)$ to complex $j$, the even-spin Sommerfeld--Watson identity reads
\begin{equation}
\sum_{j\in 2\mathbb{N}} f(j)
=
-\frac{\pi}{2}\int_{\mathcal{C}_{\rm SW}}\frac{\dd j}{2\pi\ii}\;
\frac{1+e^{\ii\pi j}}{\sin(\pi j)}\,f(j),
\label{eqS:SW_even}
\end{equation}
where $\mathcal{C}_{\rm SW}$ is the formal Sommerfeld--Watson contour encircling the even
positive integers $j=2,4,\ldots$ and excluding $j=0$, see Fig.~\ref{fig:contour_app_MB}.
The contour is oriented clockwise around the positive even-spin poles; with this orientation
the minus sign in Eq.~\eqref{eqS:SW_even} gives the positive sum of residues.  This fixed-spin
contour should not be interpreted as a numerical large right semicircle at physical $s$.
Using
$(-s)^j=s^j e^{\ii\pi j}$ for $s>0$ gives
\begin{equation}
\left(\frac{s}{\kN^2}\right)^{\!j}\!\Bigl(1+e^{\ii\pi j}\Bigr)
=
\left(\frac{s}{\kN^2}\right)^{\!j}
+
\left(-\frac{s}{\kN^2}\right)^{\!j},
\label{eqS:signature_factor}
\end{equation}
which is the universal even-signature structure in the main-text amplitude.

\subsection{Residues in holographic QCD}
In our application the residue is modeled by holographic QCD dynamics.  It is of the form
\begin{equation}
\beta_{V,j}(t)\;\longrightarrow\;\mathcal{I}_V(j)\,\hat d_j(\eta,t)\,A_g(t,j),
\label{eqS:factorization}
\end{equation}
where $\mathcal{I}_V(j)$ is the photoproduction impact factor, $A_g(t,j)$ is the spin-$j$
proton form factor, and $\hat d_j$ models the skewness dependence and reproduces the
polynomiality/$D$-term structure at $j=2$.  Substituting
Eqs.~\eqref{eqS:fixedspin}--\eqref{eqS:factorization} into
Eq.~\eqref{eqS:SW_even} yields the main-text Mellin--Barnes integral.
Refs.~\cite{Mamo:2019mka,Mamo:2021tzd} tested the fixed-$j=2$ near-threshold model and
the leading-$\jzero$/BPST-cut high-energy model separately. The direct numerical
Mellin--Barnes integration used here is the all-spin holographic QCD completion that unifies those
two effective descriptions.

\subsection{Analytic structure in the $j$ plane}
\label{secS:analyticstructure}
The BPST strong-coupling spin--dimension relation used in the main text is
\begin{equation}
\Delta(j)=2+\sqrt{2\sqrt{\lambda}\,\bigl(j-\jzero\bigr)},
\qquad
\jzero=2-\frac{2}{\sqrt{\lambda}}.
\label{eqS:Delta}
\end{equation}
The square-root branch point at $j=\jzero$ generates the BPST cut for $j\le\jzero$.
Additional singularities arise from the signature factor $1/\sin(\pi j)$ and from the
$\Gamma$-function factors in the soft-wall expressions below.
The skewness factor $\hat d_j(\eta,t)=F_j(z_S)$ also has apparent half-integer singularities
when the third hypergeometric parameter $1/2-j$ is a non-positive integer,
\begin{equation}
j=n+\frac{1}{2},
\qquad n=0,1,2,\ldots .
\label{eqS:half_integer_poles}
\end{equation}
These are kinematic/skewness singularities, not Regge singularities.  From the power-series
representation of ${}_2F_1$, the pole at $j=n+1/2$ first appears at order $z_S^{n+1}$, so
\begin{equation}
\operatorname*{Res}_{j=n+1/2}F_j(z_S)=\mathcal{O}\!\left(z_S^{n+1}\right).
\label{eqS:half_integer_residue}
\end{equation}
In the physical high-energy limit, $z_S\sim\eta^2\sim s^{-2}$ at fixed $t$, and the
corresponding contribution behaves parametrically as
\begin{equation}
s^{n+1/2}z_S^{n+1}\sim s^{-n-3/2},
\label{eqS:half_integer_suppression}
\end{equation}
up to logarithms and slowly varying $t$-dependent factors.  These kinematic singularities
therefore do not modify the BPST leading high-energy behavior.

\subsection{Contour deformation and convergence}
\label{subsecS:convergence}
Write the main-text Mellin--Barnes integrand as
\begin{equation}
\mathcal{M}_V(s,t)
=
-\frac{\pi}{2}\int_{\mathcal{C}_B}\frac{\dd j}{2\pi\ii}\;\Phi_V(j;s,t),
\qquad
\Phi_V(j;s,t)\equiv
\frac{\left(\dfrac{s}{\kN^2}\right)^j+\left(-\dfrac{s}{\kN^2}\right)^j}{\sin(\pi j)}
\;\mathcal{I}_V(j)\;\hat d_j(\eta,t)\,A_g(t,j).
\label{eqS:Phi}
\end{equation}
We take a Bromwich line
\begin{equation}
\mathcal{C}_B:\qquad j=c+\ii\nu,
\qquad \nu\in(-\infty,\infty),
\qquad \jzero<c<2,
\qquad c\notin\mathbb{Z}.
\label{eqS:Bromwich}
\end{equation}
This line is oriented upward, lies to the right of the BPST branch point, and lies to the left of
the first fixed-spin pole.  For the Reggeized holographic amplitude used in the numerical
evaluation, the relevant deformation of $\mathcal{C}_B$ is to the left, toward the BPST cut.
The associated large left arc vanishes, as checked in Fig.~\ref{fig:MB_arc_app}.  A large right
semicircle is not used in the numerical evaluation and is not expected to vanish for physical
$s/\kN^2>1$.  Deformations that cross isolated singularities are understood with the usual
indentation/residue prescription.  The fixed-spin poles $j=2,4,\ldots$ are used only as
residue/truncation diagnostics of the analytically continued integrand, not as the numerical
contour used for the resummed evaluation.

\paragraph{Large-$|\mathrm{Im}(j)|$ damping.}
With the branch convention $(-s)^j=s^j e^{\ii\pi j}$ and $j=c+\ii\nu$,
\begin{equation}
\left|\left(\frac{s}{\kN^2}\right)^j\right|=\left(\frac{s}{\kN^2}\right)^c,
\qquad
\left|\left(-\frac{s}{\kN^2}\right)^j\right|=\left(\frac{s}{\kN^2}\right)^c e^{-\pi\nu}.
\label{eqS:Regge_moduli}
\end{equation}
The relevant object is the combined even-signature factor
\begin{equation}
\Sigma(j;s)\equiv
\frac{\left(\dfrac{s}{\kN^2}\right)^j+\left(-\dfrac{s}{\kN^2}\right)^j}{\sin(\pi j)}.
\label{eqS:Sigma}
\end{equation}
It is exponentially suppressed for $\nu\to+\infty$ and bounded, up to constants and powers,
for $\nu\to-\infty$:
\begin{equation}
|\Sigma(c+\ii\nu;s)|\le C\left(\frac{s}{\kN^2}\right)^c
\begin{cases}
e^{-\pi\nu}, & \nu\to+\infty,\\[2pt]
1, & \nu\to-\infty,
\end{cases}
\label{eqS:Sigma_bound}
\end{equation}
where $C$ depends on $c$ but not on $\nu$.  The falloff in the direction where the signature
factor is only bounded is supplied by the soft-wall overlap factors.  Using
\begin{equation}
|\Gamma(x+\ii y)|
\sim
\sqrt{2\pi}\;|y|^{x-\tfrac12}\;e^{-\tfrac{\pi}{2}|y|}
\qquad (|y|\to\infty),
\label{eqS:Stirling}
\end{equation}
one obtains
\begin{equation}
|\Phi_V(c+\ii\nu;s,t)|
\le
\mathrm{poly}(|\nu|)
\begin{cases}
e^{-\alpha_+\nu}, & \nu\to+\infty,\\[2pt]
e^{-\alpha_-|\nu|}, & \nu\to-\infty,
\end{cases}
\qquad \alpha_\pm>0,
\label{eqS:Phi_bound}
\end{equation}
consistent with the numerical decay shown in Fig.~\ref{fig:MB_decay_im_app}.  Thus the
Bromwich integral is absolutely convergent.

\section{Soft-wall model input}
\label{secS:ingredients}

\subsection{Impact factor at $Q^2=0$}
\label{secS:impactfactor}
In this section we present the explicit expressions of the impact factor $\mathcal{I}_V(j)$
and the proton coupling $A_g(t,j)$ in the soft-wall model of holographic QCD.
The expressions are obtained from the analytic continuation of the spin-$j$ soft-wall
overlap integrals and are analytic in complex $j$ up to the BPST cut and isolated poles.

We parameterize the photoproduction-point impact factor as
\begin{equation}
\mathcal{I}_V(j)=
\frac{\mathcal{C}_V(j)}{\Gamma\!\left(\dfrac{j+\Delta(j)+2}{2}\right)},
\label{eqS:Ij}
\end{equation}
with
\begin{align}
\mathcal{C}_V(j)
&=\frac{N_V}{2}\left(\frac{j+\Delta(j)}{2}\right)
\frac{\Gamma^2\!\left(\frac{j+\Delta(j)}{2}\right)}{\Delta(j)}
\left(\frac{\kV}{\kN}\right)^{j+10-3\Delta(j)}.
\label{eqS:Cj}
\end{align}
The effective normalization $N_V$ absorbs the photon--vector-meson overlap, bulk wave-function
conventions, and the remaining overall amplitude normalization.  It also absorbs residual
channel dependence associated with the common heavy-quarkonium soft-wall scale used below.
In the numerical work the $J/\psi$ normalization is fitted together with $\lambda$; the
independently specified $\Upsilon$ normalization used for the $\Upsilon$ curve is given in
Table~\ref{tab:params}.

\subsection{Spin-$j$ proton form factor}
\label{secS:protonff}
We write
\begin{align}
A_g(t,j)
&=
A_g(0)\,2^{-2+\Delta}\,
\Gamma\!\left(a_K+\frac{\Delta}{2}\right)
\Bigl[\,\mathcal{T}_0(t,j)+\mathcal{T}_1(t,j)\Bigr],
\label{eqS:Atj}
\\
a_K&\equiv -\frac{t}{8\kN^2},
\qquad
\Delta\equiv \Delta(j),
\nonumber
\end{align}
with
\begin{align}
\mathcal{F}_n(t,j)
&\equiv
{}_2\tilde{F}_{1}\!\Bigl(
2+a_K-\frac{\Delta}{2},
\frac{j}{2}+\tau-\frac{\Delta}{2}+n;
 a_K+\frac{j}{2}+\tau+n;
-1
\Bigr),
\quad n=0,1,
\label{eqS:Fn}
\\[2pt]
\mathcal{T}_0(t,j)
&\equiv
\frac{2^{1-\Delta}(\tau-1)}{\Gamma(\tau)}\,
\Gamma\!\left(\frac{j}{2}+\tau-\frac{\Delta}{2}\right)
\Gamma\!\left(\frac{j}{2}+\tau+\frac{\Delta}{2}-2\right)\,
\mathcal{F}_0(t,j),
\label{eqS:T0}
\\[2pt]
\mathcal{T}_1(t,j)
&\equiv
\frac{2^{1-\Delta}}{\Gamma(\tau)}\,
\Gamma\!\left(\frac{j}{2}+\tau-\frac{\Delta}{2}+1\right)
\Gamma\!\left(\frac{j}{2}+\tau+\frac{\Delta}{2}-1\right)\,
\mathcal{F}_1(t,j),
\label{eqS:T1}
\end{align}
where ${}_2\tilde{F}_1$ is the regularized Gauss hypergeometric function.  The parameter
$\tau$ is the effective twist of the baryonic interpolating operator; in the numerical
implementation we use $\tau=3$.

\section{Skewness in holographic QCD}

\subsection{Skewness dependence and $D$-term}
\label{secS:dterm}
In this section we describe the treatment of the skewness (longitudinal momentum transfer)
in heavy quarkonium photoproduction in the holographic QCD formulation. The skewness dependence
of the holographic proton coupling is derived by matching with the polynomiality condition
of the GPD moments.

The photoproduction process is characterized by the skewness
\begin{equation}
\eta = \frac{\Delta\cdot \tilde{q}}{\tilde{p}\cdot \tilde{q}}
=
\frac{M_{V}^2}{2(s-m_N^2)-M_{V}^2+t}.
\label{eqS:eta}
\end{equation}
We use the following holographic model prescription for the skewness dependence of the
complex-$j$ integrand:
\begin{equation}
A_g(t,j)+D_\eta(t,j)=\hat d_j(\eta,t)\,A_g(t,j),
\label{eqS:Aj_plus_D}
\end{equation}
where $\hat d_j$ is the hypergeometric factor in the main text,
\begin{equation}
\hat d_j(\eta,t)
=
{}_2F_{1}\!\left(
-\frac{j}{2},
\frac{1-j}{2};
\frac{1}{2}-j;
-\frac{4m_N^2\eta^2}{t-m_S^2}
\right).
\label{eqS:dhat}
\end{equation}
For positive even $j$, Eq.~\eqref{eqS:dhat} terminates to a polynomial in $\eta^2$ and
serves as the holographic polynomiality input used in the complex-$j$ integrand.
At $j=2$ one recovers the conventional $D$-term structure,
\begin{equation}
D_\eta(t,2)=\eta^2D_g(t,2)
=\eta^2\frac{(4m_N^2/3)}{t-m_S^2}A_g(t,2),
\end{equation}
with $D_g(t)=4C_g(t)$ in standard gluonic GFF notation.  Therefore
\begin{equation}
D_g(t,2)=\frac{4m_N^2}{3}\frac{A_g(t,2)}{t-m_S^2}.
\label{D2}
\end{equation}
Away from positive even spin, the same closed holographic expression in
Eq.~\eqref{eqS:dhat} is used as the complex-$j$ input to the Mellin--Barnes amplitude;
the $j=2$ limit fixes the normalization to the standard GFF polynomiality structure.

\begin{figure*}[t]
\centering
\subfloat[$\eta(E_\gamma,t_{\min})$]{%
  \includegraphics[width=.32\linewidth]{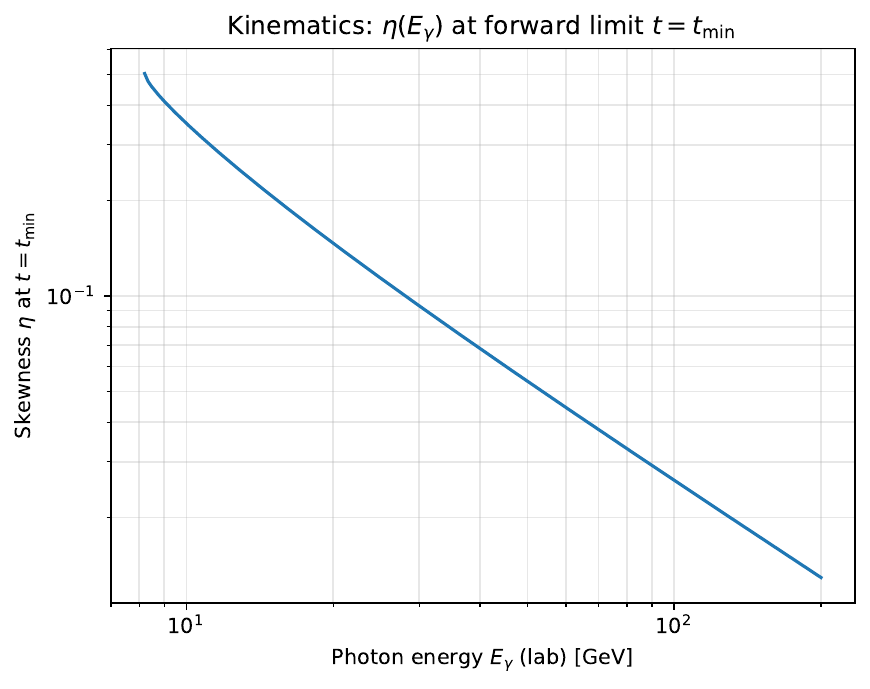}}
\hfill
\subfloat[$-t_{\min}(E_\gamma)$]{%
  \includegraphics[width=.32\linewidth]{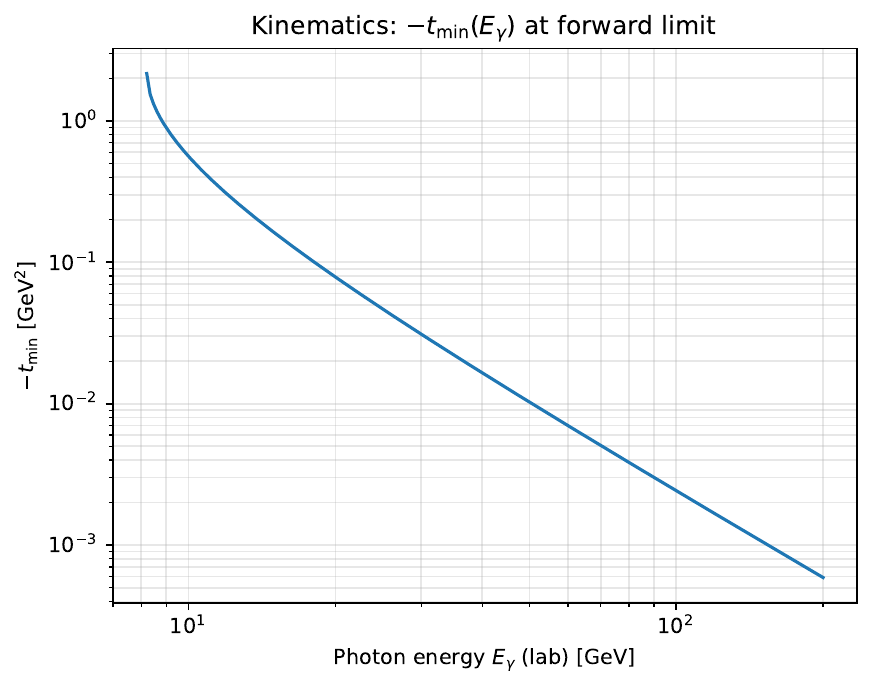}}
\hfill
\subfloat[$z_S(E_\gamma,t_{\min})$]{%
  \includegraphics[width=.32\linewidth]{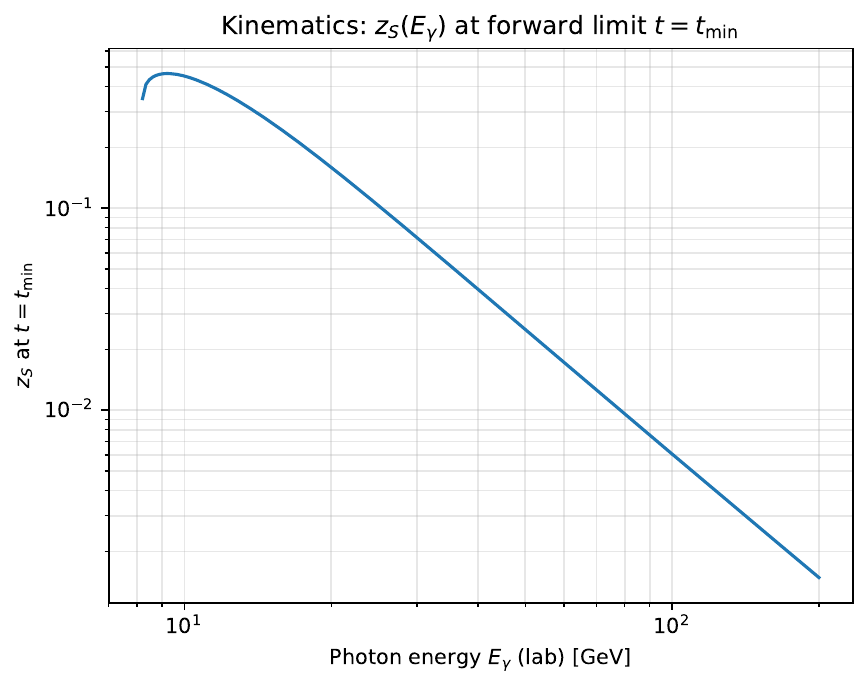}}
\caption{Photoproduction kinematic path at the forward endpoint $t=t_{\min}(s)$ using the
same conventions and parameters as the convergence tests. The skewness $\eta$ decreases with
lab photon energy, $-t_{\min}$ approaches zero, and the hypergeometric argument $z_S$ remains
below unity and decreases rapidly at high energy.}
\label{fig:photoproduction_kinematics}
\end{figure*}

\paragraph{Argument in the physical region.}
For spacelike momentum transfer, $t=-K^2<0$, the argument of the hypergeometric factor is
\begin{equation}
z_S(t)\equiv -\frac{4m_N^2\eta^2}{t-m_S^2}
=\frac{4m_N^2\eta^2}{K^2+m_S^2}\ge0.
\label{eqS:zS}
\end{equation}
The exact two-body physical-region condition is
\begin{equation}
\eta^2\le
\frac{-M_V^4 t}
{(4m_N^2-t)(M_V^2-t)^2},
\label{eqS:eta_physical_bound}
\end{equation}
where equality holds at the collinear kinematic boundaries. This follows from the
Gram-determinant condition, or equivalently from the non-positive norm of the component
of $\tilde q$ transverse to $\tilde p$ and $\Delta$. It then follows that
\begin{equation}
0\le z_S(t)\le
\frac{4m_N^2M_V^4(-t)}
{(4m_N^2-t)(M_V^2-t)^2(m_S^2-t)}
<1.
\label{eqS:zS_physical_bound}
\end{equation}
The final inequality is immediate for $t=-K^2<0$ and $m_S^2>0$, because
$(4m_N^2+K^2)>4m_N^2$, $(M_V^2+K^2)^2>M_V^4$, and
$(m_S^2+K^2)>K^2$. Thus $0\le z_S<1$ throughout the physical
photoproduction region.
Along the actual photoproduction kinematic path, $\eta=\eta(s,t)$ varies with both $s$
and $t$. Evaluated at the forward endpoint $t=t_{\min}(s)$, the physical path has
decreasing $\eta$ and decreasing $z_S$ with increasing lab photon energy, as shown in
Fig.~\ref{fig:photoproduction_kinematics}.

\subsection{Large-$j$ behavior from WKB approximation}
The large-$j$ behavior of the skewness factor can be derived in a semiclassical WKB-type
approximation. Define
\begin{equation}
\mathcal{D}_j(z)\equiv {}_2F_1\!\left(-\frac{j}{2},\frac{1-j}{2};\frac{1}{2}-j;z\right),
\qquad \hat d_j(\eta,t)=\mathcal{D}_j(z_S).
\label{eqS:Fjdef}
\end{equation}
$\mathcal{D}_j$ satisfies the hypergeometric differential equation
\begin{align}
z(1-z)\mathcal{D}_j''(z)
+\Bigl[\Bigl(\frac{1}{2}-j\Bigr)-\Bigl(\frac{3}{2}-j\Bigr)z\Bigr]\mathcal{D}_j'(z)
-\frac{j(j-1)}{4}\,\mathcal{D}_j(z)=0.
\label{eqS:FjODE}
\end{align}
Using a WKB ansatz $\mathcal{D}_j(z)=A(z)\exp\!\bigl(j\phi(z)\bigr)\,[1+\Order{1/j}]$ gives,
for $0\le z<1$ and the branch selected by $D_j(0)=1$,
\begin{equation}
\phi(z)=\ln\!\left(\frac{1+\sqrt{1-z}}{2}\right),
\qquad
A(z)=\sqrt{\frac{1+\sqrt{1-z}}{2\sqrt{1-z}}}.
\label{eqS:WKBphiA}
\end{equation}
Therefore
\begin{align}
\mathcal{D}_j(z)
&=
\sqrt{\frac{1+\sqrt{1-z}}{2\sqrt{1-z}}}\,
\left(\frac{1+\sqrt{1-z}}{2}\right)^j
\Bigl[1+\Order{1/j}\Bigr],
\qquad (0\le z<1),
\label{eqS:FjWKB}
\end{align}
which implies exponential damping at fixed $z\in(0,1)$ as $j\to\infty$.  For $z\ll1$,
\begin{equation}
\mathcal{D}_j(z)\sim\exp\!\left(-\frac{jz}{4}\right),
\qquad (z\ll1),
\label{eqS:Fjsmallz}
\end{equation}
up to an order-one prefactor.

\section{Input parameters and global fit}
\label{secS:globalfit}

\subsection{Input parameters}
In this section we describe the determination of the input parameters of the holographic QCD model
and the procedure of the fit to the heavy quarkonium cross section data.

\begin{figure*}[t]
\centering
\subfloat[$A_g(t,2)$\label{fig_Alatticefit}]{%
  \includegraphics[height=5.5cm,width=.46\linewidth]{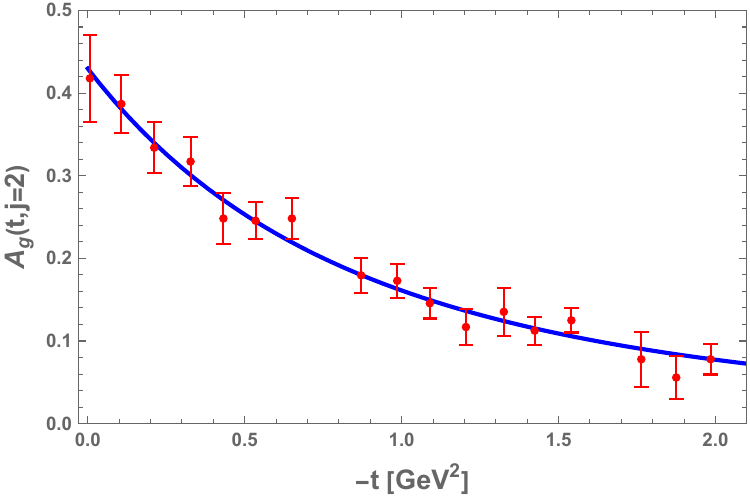}%
}\hfill
\subfloat[$D_g(t,2)$\label{fig_Dlatticefit}]{%
  \includegraphics[height=5.5cm,width=.46\linewidth]{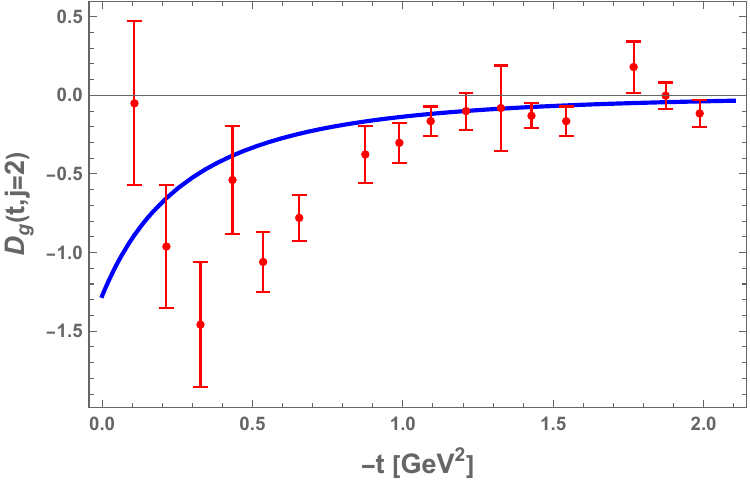}%
}
\caption{Lattice-QCD normalization of the spin-$2$ gluonic input.  The lattice data are the
$\overline{\mathrm{MS}}$ gluon EMT form factors at $\mu=2~\mathrm{GeV}$.  (a) The gluonic form
factor $A_g(t,2)$ from lattice QCD \cite{Pefkou:2021fni} and the holographic-QCD fit using
Eq.~\eqref{eqS:Atj}, which fixes $A_g(0)=0.430$ and $\kN=0.388~\mathrm{GeV}$.  (b) The gluonic
$D$-term form factor from lattice QCD \cite{Pefkou:2021fni} and the holographic-QCD fit using
Eq.~\eqref{D2}; with $A_g(0)$ and $\kN$ fixed from panel (a), this determines
$m_S=0.630~\mathrm{GeV}$.}
\label{fig_A-D_latticefit}
\end{figure*}

The proton-side spin-$2$ gluonic input is fixed independently of the photoproduction data.
The lattice-QCD gluon EMT form factors used here are the $\overline{\mathrm{MS}}$-renormalized
gluon contributions at $\mu=2~\mathrm{GeV}$.  We fit the lattice-QCD result for the gluonic
form factor $A_g(t,2)$ \cite{Pefkou:2021fni} with Eq.~\eqref{eqS:Atj}, which fixes
$A_g(0)=0.430$ and $\kN=0.388~\mathrm{GeV}$.  With these values held fixed, the lattice-QCD
result for the $D$-term is fitted with Eq.~\eqref{D2}, giving $m_S=0.630~\mathrm{GeV}$.
These lattice/holographic-QCD fits are shown in Fig.~\ref{fig_A-D_latticefit}.  The twist
parameter is fixed as $\tau=3$ by leading baryon power counting.

\subsection{Total cross section fit}
After these inputs are fixed, the photoproduction fit determines only the 't Hooft coupling
$\lambda$, which fixes $\jzero=2-2/\sqrt{\lambda}$, and the $J/\psi$ normalization
$N_{J/\psi}$ in the impact factor.  We minimize
\begin{equation}
\chi^2(\lambda,N_{J/\psi})
=
\sum_{i\in \mathcal{D}_{J/\psi}^{\rm glob}}
\frac{\left[
\sigma_i^{\rm exp}
-
\sigma^{\rm th}(E_{\gamma,i};\lambda,N_{J/\psi})
\right]^2}{\delta\sigma_i^2},
\label{eqS:chi2_global}
\end{equation}
where $\sigma^{\rm th}$ is obtained from Eq.~\eqref{eqS:sigma_total}.  The fit set
$\mathcal{D}_{J/\psi}^{\rm glob}$ is the full threshold-to-HERA world $J/\psi$
total-cross-section data set shown in the main text, including the modern JLab near-threshold
measurements together with the fixed-target and HERA data
\cite{GlueX:2019mkq,GlueX:2023pev,007:2026dow,CLAS:2026lls,%
Gittelman:1975ix,Camerini:1975cy,Barate:1986fq,Binkley:1981kv,E687:1993hlm,%
H1:2000kis,ZEUS:2002wfj,H1:2005dtp,H1:2013okq}.
All threshold-to-HERA $J/\psi$ total-cross-section data shown in the main-text fit figure are
included in the global fit. The $\Upsilon$ points are used only as an external channel check,
and the ultra-high-energy UPC points are not included in the fit; they are shown as a qualitative
indication consistent with the expected onset of eikonal/unitarity corrections beyond the
single-cut approximation.
The minimization gives
\begin{equation}
\lambda=8.13,
\qquad
N_{J/\psi}^{2}=5.27\times10^{-5}\left[\text{GeV}^2\,\text{nb}\right],
\qquad
\chi^2/\mathrm{d.o.f.}\simeq 2.27.
\label{fit_results}
\end{equation}
The value of $\chi^2/\mathrm{d.o.f.}$ should be interpreted in the context of the assembled
world data set.  The global fit combines measurements from different experiments and analyses,
some of which have overlapping or nearby $E_\gamma$ coverage but mutually displaced central
values and normalization systematics.  Since Eq.~\eqref{eqS:chi2_global} does not introduce
experiment-dependent floating normalizations or correlated normalization nuisance parameters,
these inter-data-set tensions contribute directly to the quoted $\chi^2$.  The resulting
curve should therefore be viewed as the best common smooth energy dependence obtained in the
minimal two-parameter single-cut model.  In overlapping kinematic regions it follows one of the
coexisting determinations rather than forcing a compromise through mutually displaced points.
A dedicated fit-set and correlated-systematics study is required for a more statistical
decomposition of the residual $\chi^2$.
The same Regge/proton parameter set is then used for the $\Upsilon$ curve in the main text,
together with the independently specified channel normalization
$N_{\Upsilon}^{2}=9.98\times10^{-8}\left[\text{GeV}^2\,\text{nb}\right]$.  This $\Upsilon$ comparison is not parameter-free; it
tests whether the same Regge/proton sector and energy dependence can accommodate the
$\Upsilon$ data after fixing the channel normalization.  The complete parameter set is
summarized in Table~\ref{tab:params}.

\begin{table}[t]
\centering
\small
\renewcommand{\arraystretch}{1.18}
\begin{tabular}{@{}lll@{}}
\toprule
\tpar{2.4cm}{Symbol} & \tpar{2.6cm}{Value} & \tpar{8.8cm}{How fixed} \\
\midrule
\multicolumn{3}{@{}l@{}}{\textit{Photoproduction fit and channel parameters}}\\
\midrule
$\lambda$ & $8.13$ &
\tpar{8.8cm}{Fitted by $\chi^2$ minimization to the full threshold-to-HERA
$\gamma p\to J/\psi p$ total-cross-section data; fixes
$\jzero=2-2/\sqrt{\lambda}\simeq1.30$.} \\[3pt]

$N_{J/\psi}^{2}$ & $5.27\times10^{-5}\left[\text{GeV}^2~\text{nb}\right]$ &
\tpar{8.8cm}{$J/\psi$ impact-factor normalization fitted to the same global
$J/\psi$ data set.} \\[3pt]

$N_{\Upsilon}^{2}$ & $9.98\times10^{-8}\left[\text{GeV}^2\,\text{nb}\right]$ &
\tpar{8.8cm}{Independently specified $\Upsilon$ impact-factor normalization used for the
$\Upsilon$ curve shown in the main text; the $\Upsilon$ data are not included in the
$J/\psi$ fit.} \\[3pt]

\midrule
\multicolumn{3}{@{}l@{}}{\textit{Gluonic-GFF inputs fixed from lattice-QCD/holographic-QCD fits}}\\
\midrule
$A_g(0)$ & $0.430$ &
\tpar{8.8cm}{Fixed by fitting the $\overline{\mathrm{MS}}$, $\mu=2~\mathrm{GeV}$ lattice-QCD
$A_g(t,2)$ form factor with Eq.~\eqref{eqS:Atj}; see Fig.~\ref{fig_Alatticefit}.} \\[3pt]

$\kN$ & $0.388~\mathrm{GeV}$ &
\tpar{8.8cm}{Fixed by the same lattice-QCD/holographic-QCD $A_g(t,2)$ fit; controls the proton-side
soft-wall scale.} \\[3pt]

$m_S$ & $0.630~\mathrm{GeV}$ &
\tpar{8.8cm}{Fixed by fitting the lattice-QCD $D_g(t,2)$ form factor with Eq.~\eqref{D2},
using $A_g(0)$ and $\kN$ from the $A_g(t,2)$ fit; see Fig.~\ref{fig_Dlatticefit}.} \\[3pt]

$\tau$ & $3$ &
\tpar{8.8cm}{Fixed by leading three-quark/baryon power counting.} \\[3pt]

\midrule
\multicolumn{3}{@{}l@{}}{\textit{Fixed vector-meson input}}\\
\midrule
$\kV\equiv\kappa_{J/\psi}\equiv\kappa_{\Upsilon}$ & $1.038~\mathrm{GeV}$ &
\tpar{8.8cm}{Fixed from the $Q^2$ dependence of $J/\psi$ electroproduction in
Ref.~\cite{Mamo:2021tzd}.  The equality
$\kappa_{J/\psi}=\kappa_{\Upsilon}$ is a deliberate heavy-quarkonium universality
approximation and practical simplification; residual channel dependence is absorbed in
the independent normalization $N_{\Upsilon}$.} \\
\bottomrule
\end{tabular}
\caption{Parameter set used in the numerical evaluation of
$\gamma p\to Vp$, with $V=J/\psi,\Upsilon$.  Only $\lambda$ and $N_{J/\psi}$ are fitted to the
full global $J/\psi$ total-cross-section data; $N_{\Upsilon}$ is an independently specified
channel normalization.  The proton-side gluonic-GFF inputs are fixed before the photoproduction
fit by the lattice-QCD/holographic-QCD comparison in Fig.~\ref{fig_A-D_latticefit}.}
\label{tab:params}
\end{table}

\subsection{Differential cross section prediction}
\label{secS:differential_prediction}

The total-cross-section fit in Eq.~\eqref{eqS:chi2_global} fixes $\lambda$ and
$N_{J/\psi}$.  Once these two parameters are fixed, the differential cross section in
Eq.~\eqref{eqS:dsdt} is a prediction of the same resummed amplitude.  No differential
cross-section points are included in the $\chi^2$ minimization, and no additional
$t$-slope, bin-dependent normalization, or shape parameter is refitted.

\begin{figure*}[t]
\centering
\subfloat[GlueX bin, $E_\gamma\simeq10.96~\mathrm{GeV}$\label{fig_Diff_CS_GlueX23}]{%
  \includegraphics[height=5.5cm,width=.46\linewidth]{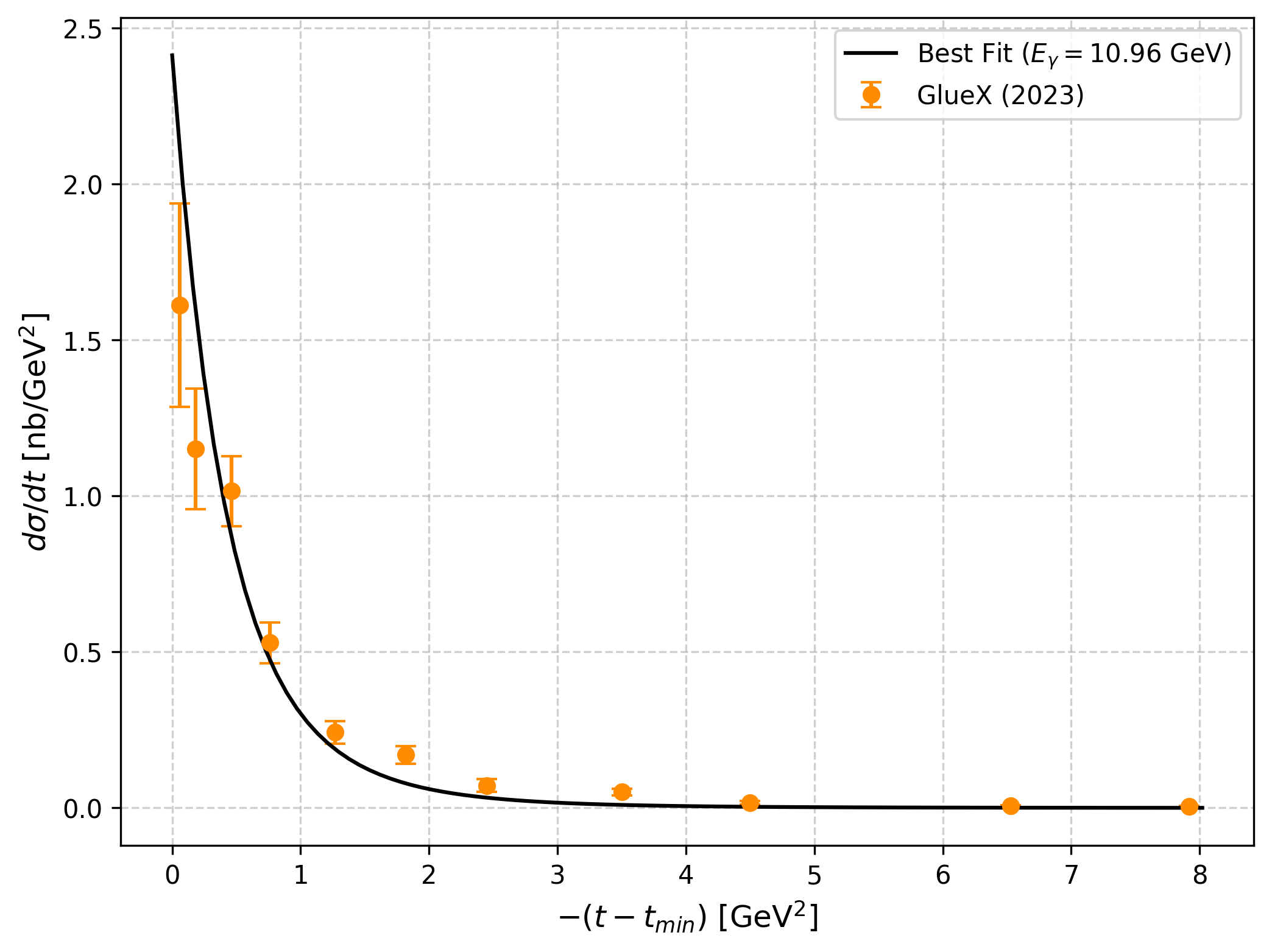}%
}\hfill
\subfloat[CLAS bin, $E_\gamma\simeq10.3~\mathrm{GeV}$\label{fig_Diff_CS_CLAS26}]{%
  \includegraphics[height=5.5cm,width=.46\linewidth]{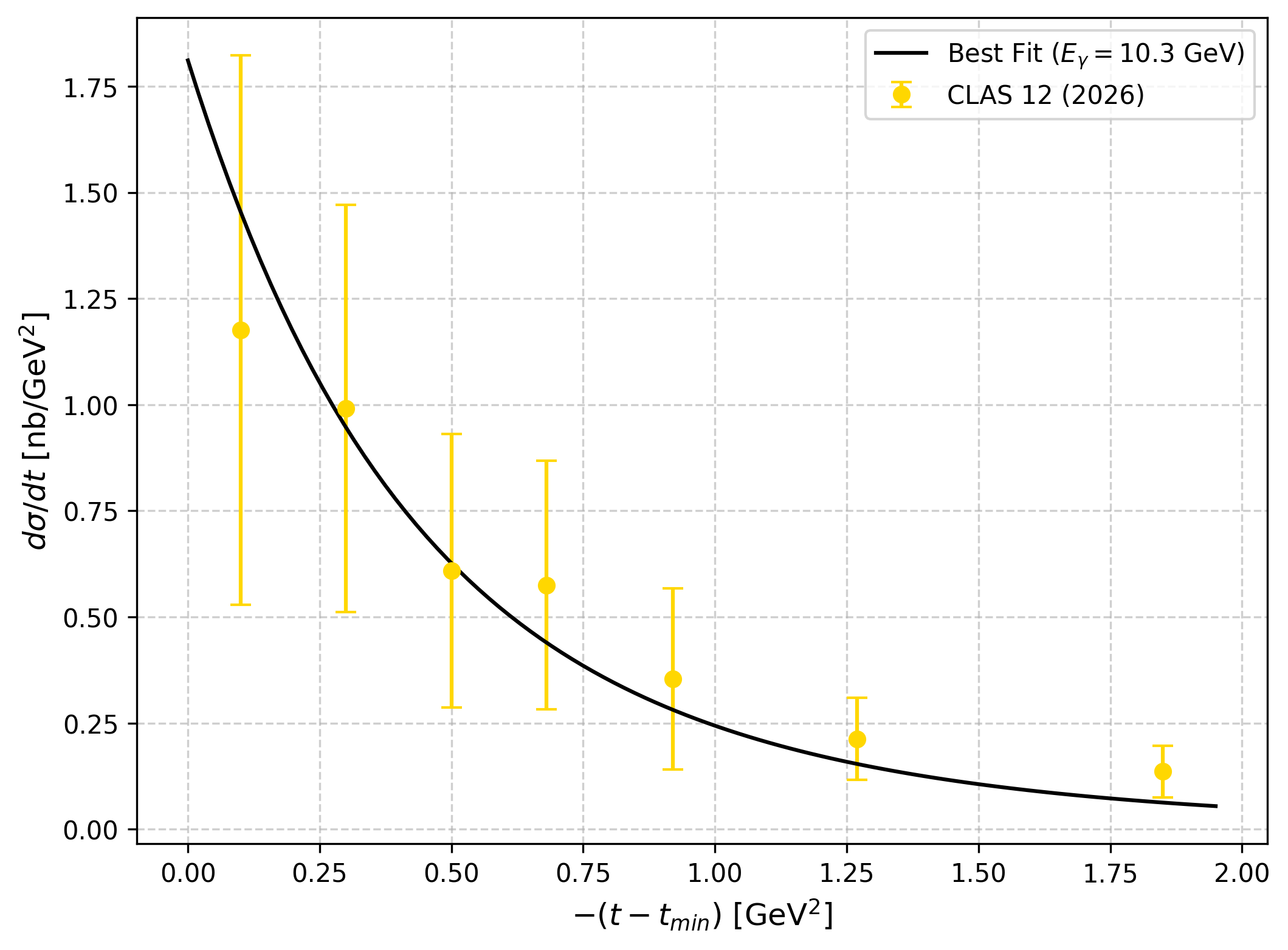}%
}
\caption{Prediction-level comparison with representative near-threshold differential
cross-section data.  The curves use the same parameter set in Table~\ref{tab:params},
obtained from the total $J/\psi$ cross-section fit; the differential data shown here are not
included in the fit.  The horizontal variable is the shifted momentum transfer
$-(t-t_{\min})$, so the forward endpoint is at the origin.}
\label{fig:diff_cs_jlab}
\end{figure*}

Figure~\ref{fig:diff_cs_jlab} compares this prediction with representative near-threshold
differential cross-section bins from GlueX and CLAS.  The resulting agreement provides a
nontrivial check of the predicted $t$ dependence generated by the lattice-fixed proton input,
the skewness factor, and the resummed complex-$j$ amplitude.  A full differential analysis,
including all available JLab bins from CLAS, GlueX, and related measurements, as well as
high-energy HERA differential cross-section data, is left for follow-up work.

\section{Numerical evaluation of Mellin-Barnes integral}
\label{secS:numerics}

\subsection{Bromwich contour}
In this section we discuss the choice of the integration contour in the Mellin-Barnes integral
and the numerical evaluation of the integral.

A stable implementation evaluates the main-text Mellin--Barnes integral by parameterizing
the Bromwich line as $j=c+\ii\nu$ and integrating over $\nu\in(-\infty,\infty)$ with adaptive
quadrature.  In practice one truncates to $|\nu|\le\nu_{\max}$ and checks stability with
increasing $\nu_{\max}$; the combined signature factor and the soft-wall $\Gamma$-function
suppression make this efficient.  We have verified numerical stability under variations of
$c$ within the allowed window $\jzero<c<2$ and under increases of $\nu_{\max}$.  One must
choose a principal branch for the square root in $\Delta(j)$, keep $\jzero<c<2$ while avoiding
integers and half-integer skewness poles, and evaluate special functions with sufficient
precision for large complex arguments.

\paragraph{Typical choice.}
A representative choice used in convergence tests is $c=1.6$.  For the full global fit,
$\lambda=8.13$, so $\jzero\simeq1.30$ and this contour lies to the right of the BPST branch
point and to the left of the first positive fixed-spin pole at $j=2$.  The parameter set used
for the $J/\psi$ convergence plots is
\begin{align}
& \kappa_V=1.038~\mathrm{GeV},\quad
\kappa_N=0.388~\mathrm{GeV},\quad
\lambda=8.13,
\quad
N_{J/\psi}^{2}=5.27\times10^{-5}\left[\text{GeV}^2\,\text{nb}\right],
\nonumber \\
& A_g(0)=0.430,
\quad
m_S=0.630~\mathrm{GeV}.
\end{align}
For the energy we choose two representative lab energies,
$E_\gamma=9~\mathrm{GeV}$ and $E_\gamma=100~\mathrm{GeV}$, and set $t=t_{\min}(s)$.

\subsection{Diagnostic figures}
\label{secS:figs}

\begin{figure}[t]
  \centering
  \includegraphics[width=.6\textwidth]{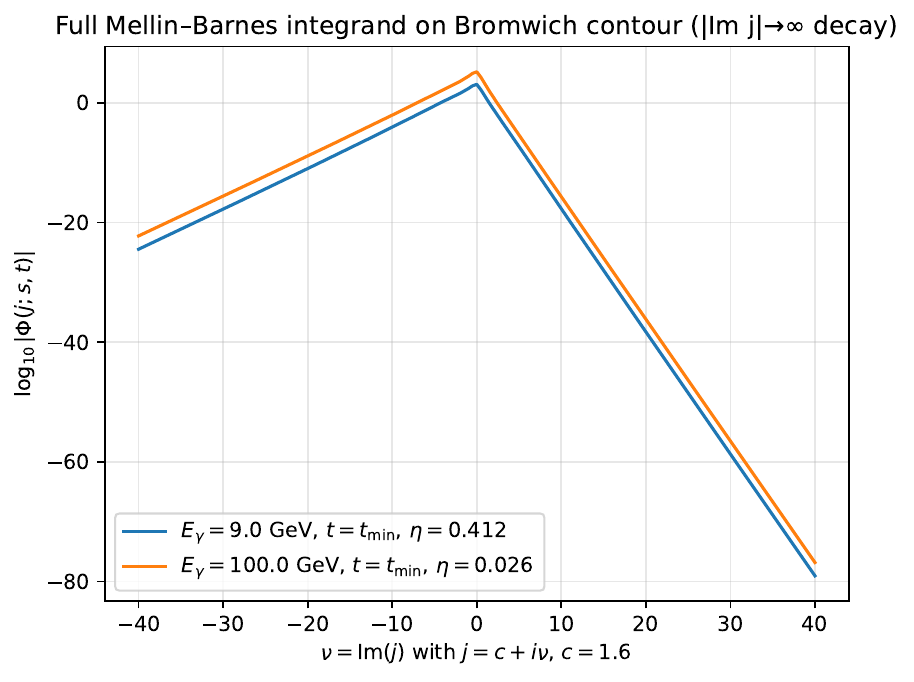}
  \caption{Magnitude of the Mellin--Barnes integrand on the Bromwich contour
  $j=c+\ii\nu$ with $c=1.6$.  The falloff for $|\nu|\to\infty$ follows from the combined
even-signature factor and the Stirling suppression of the soft-wall $\Gamma$-function factors.}
  \label{fig:MB_decay_im_app}
\end{figure}

\begin{figure}
  \centering
  \includegraphics[width=.6\textwidth]{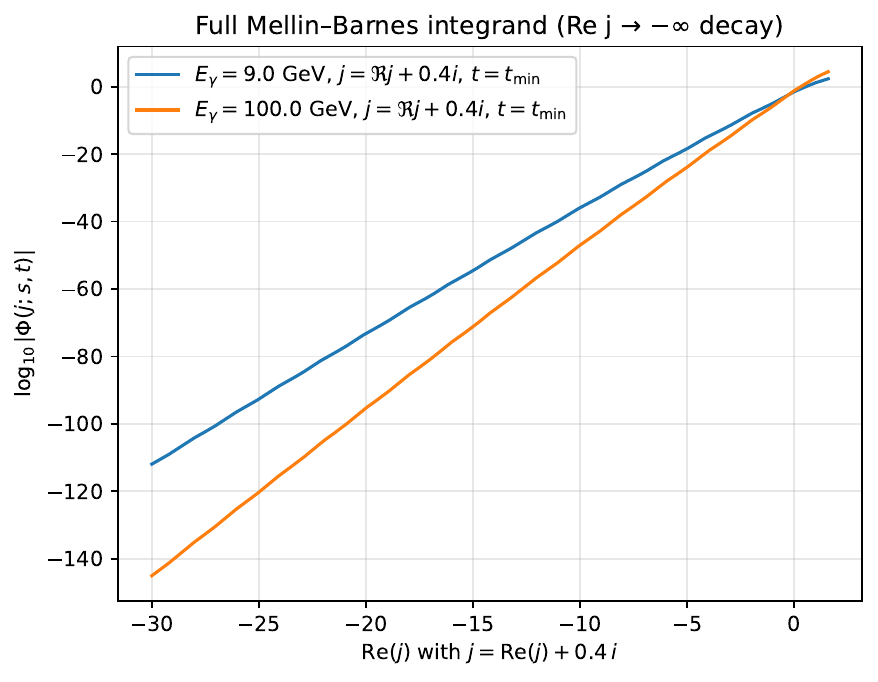}
  \caption{Left-half-plane suppression of the Mellin--Barnes integrand along
  $j=\mathrm{Re}(j)+0.4\,\ii$.  This behavior is the convergence check relevant for deforming
the Bromwich contour toward the BPST cut.}
  \label{fig:MB_decay_re_app}
\end{figure}

\begin{figure}
  \centering
  \includegraphics[width=.6\textwidth]{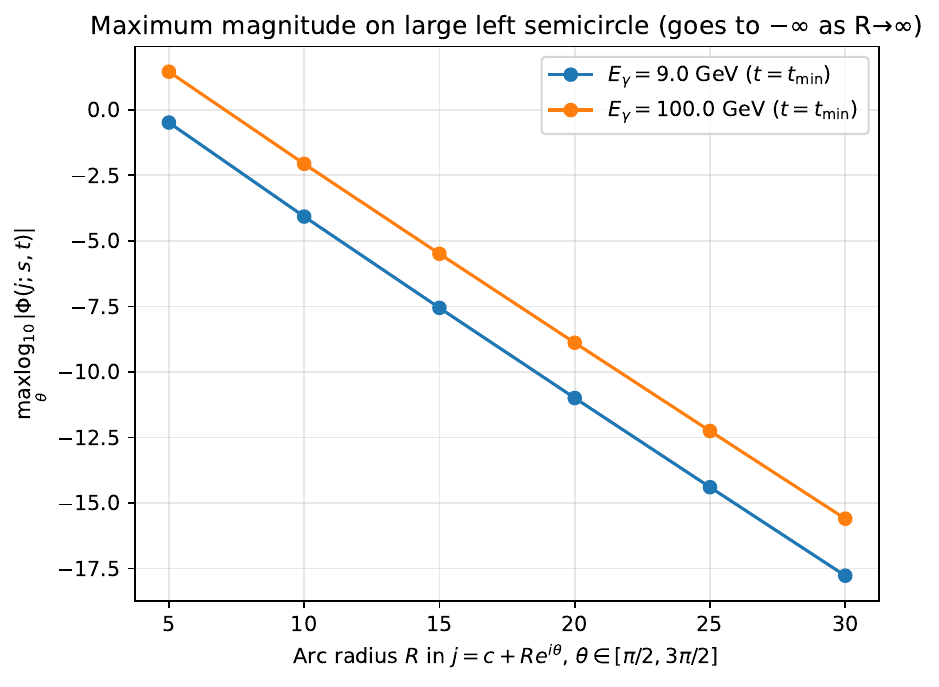}
  \caption{Large-left-arc convergence check for $j=c+Re^{\ii\theta}$,
  $\theta\in[\pi/2,3\pi/2]$.  The decrease of $\max_\theta|\Phi_V|$ with $R$ confirms
numerically that the large arc does not contribute in the BPST-cut deformation.}
  \label{fig:MB_arc_app}
\end{figure}

Figure~\ref{fig:MB_decay_im_app} plots $\log_{10}|\Phi_V(j;s,t)|$ on the Bromwich line
$j=c+\ii\nu$ with $c=1.6$.  The integrand decreases rapidly for $|\nu|\to\infty$, confirming
the bound and the absolute convergence of the $\nu$ integral.

Figure~\ref{fig:MB_decay_re_app} shows the decay for $\mathrm{Re}(j)\to-\infty$ along
a line $j=\mathrm{Re}(j)+0.4\,\ii$.  The small imaginary part avoids isolated poles on the
real axis.  This suppression is the diagnostic relevant for deforming the Bromwich contour
toward the BPST cut.

Finally, Fig.~\ref{fig:MB_arc_app} provides a direct numerical Jordan-type check.  For the left
semicircle $j=c+Re^{\ii\theta}$, $\theta\in[\pi/2,3\pi/2]$, it plots
$\max_\theta\log_{10}|\Phi_V|$ as a function of radius $R$.  The decrease with $R$ confirms
that the large left arc does not contribute in the BPST-cut deformation.

\end{document}